\documentclass[a4paper,twoside,12pt]{article}
\input{obsjkt.sty}

\newcommand{\Msun}{\ensuremath{~{\rm M}_\odot}}                   
\newcommand{\Rsun}{\ensuremath{~{\rm R}_\odot}}                   
\newcommand{\rhosun}{\ensuremath{~\rho_\odot}}                    
\newcommand{\Teff}{\ensuremath{T_{\rm eff}}}                      
\newcommand{\FeH}{\ensuremath{\rm [Fe/H]}}                        
\newcommand{\logg}{\ensuremath{\log g}}                           
\newcommand{\EBV}{\ensuremath{E(B\!-\!V)}}                        
\newcommand{\Grp}{\ensuremath{G_{\rm RP}}}                        
\newcommand{\degr}{\ensuremath{^\circ}}                           
\renewcommand{\kms}{~km~s$^{-1}$}                                 
\newcommand{\chisq}{\ensuremath{\chi^{\,2}}}                      
\newcommand{\chir}{\ensuremath{\chi_\nu^{\,2}}}                   
\newcommand{\etal}{\textit{et al.}}                               
\newcommand{\kepler}{\textit{Kepler}}                             
\newcommand{\hip}{\textit{Hipparcos}}                             
\newcommand{\gaia}{\textit{Gaia}}                                 
\newcommand{\targ}{BS~Dra}
\newcommand{\targfull}{BS~Draconis}

\newcommand{\reff}[1]{{#1}}

\newcommand{\Msunnom}{\hbox{$\mathcal{M}^{\rm N}_\odot$}}
\newcommand{\Rsunnom}{\hbox{$\mathcal{R}^{\rm N}_\odot$}}
\newcommand{\Lsunnom}{\hbox{$\mathcal{L}^{\rm N}_\odot$}}

\usepackage{rotating}

\begin{document} 

\OBSheader{Rediscussion of eclipsing binaries: \targ}{J.\ Southworth}{2026 February}

\OBStitle{Rediscussion of eclipsing binaries. Paper XXIX. \\ The F-type twin system BS Draconis}

\OBSauth{John Southworth}

\OBSinstone{Astrophysics Group, Keele University, Staffordshire, ST5 5BG, UK}


\OBSabstract{We present an analysis of \targ, a detached eclipsing binary containing two almost-identical F3\,V stars in a 3.36-d circular orbit, based on 40 sectors of observations from the Transiting Exoplanet Survey Satellite (TESS) and published spectroscopic results. We measure masses of $1.305 \pm 0.015$\Msun\ and $1.284 \pm 0.017$\Msun, and radii of $1.409 \pm 0.006$\Rsun\ and $1.400 \pm 0.006$\Rsun, for the two components. The high quality of the TESS data allow -- for the first time -- a definitive identification of the primary eclipse, which is 0.007\,mag deeper than the secondary. The primary star is the hotter, larger and more massive of the two\reff{: the ratios of the radii and surface brightnesses are both slightly but significantly below unity.} We find a distance concordant with the \gaia\ DR3 parallax and, by comparison to theoretical models, an age of $1600 \pm 300$~Myr and a slightly sub-solar chemical composition. Our mean times of primary eclipse, each representing all eclipses in one sector, have a scatter of only 0.37~s around a linear ephemeris: \targ\ may be useful as a celestial clock.}


\section*{Introduction}

Detached eclipsing binaries (dEBs) are a foundational source of measurements of the physical properties of normal stars \cite{Andersen91aarv,Torres++10aarv}. Their properties can be determined using only observational data, geometric models, and celestial mechanics. Photometric and spectroscopic data of high quality permit the measurement of mass and radius in ``well-behaved'' dEBs to precisions of better than 0.2\% \cite{Maxted+20mn}. The Detached Eclipsing Binary Catalogue \cite{Me15aspc} (DEBCat\footnote{\texttt{https://www.astro.keele.ac.uk/jkt/debcat/}}) currently lists 367 dEBs with mass and radius measurements to approximately 2\% or better, and the current series of papers \cite{Me20obs} aims to increase this number using new photometry from space missions \cite{Me21univ} such as TESS \cite{Ricker+15jatis} and \kepler\ \cite{Borucki16rpph}.

In this work we present an analysis of \targfull\ (Table~\ref{tab:info}), which consists of two F3\,V stars so similar that previous work has been unable to definitively establish which is the primary component. It is sited in the northern continuous viewing zone of TESS so has been extensively observed throughout its mission. \reff{Below we are able to identify with certainty which is the primary eclipse, as it is 0.007~mag deeper than the secondary eclipse. We refer to the primary component (eclipsed at primary eclipse) as star~A and its companion as star~B.}


\section*{\targfull}

\begin{table}[t]
\caption{\em Basic information on \targfull. 
The $BV$ magnitudes are each the mean of 91 individual measurements \cite{Hog+00aa} distributed approximately randomly in orbital phase. 
The $JHK_s$ magnitudes from 2MASS \cite{Cutri+03book} were obtained at an orbital phase of 0.885. \label{tab:info}}
\centering
\begin{tabular}{lll}
{\em Property}                            & {\em Value}                 & {\em Reference}                      \\[3pt]
Right ascension (J2000)                   & 14 22 49.698                & \citenum{Gaia23aa}                   \\
Declination (J2000)                       & +14 56 20.14                & \citenum{Gaia23aa}                   \\
Henry Draper designation                  & HD 190020                   & \citenum{CannonPickering20anhar}     \\
\textit{Hipparcos} designation            & HIP 98118                   & \citenum{ESA97}                      \\
\textit{Tycho} designation                & TYC 4457-2347-1             & \citenum{Hog+00aa}                   \\
\textit{Gaia} DR3 designation             & 2287962824139508224         & \citenum{Gaia21aa}                   \\
\textit{Gaia} DR3 parallax (mas)          & $5.1193 \pm 0.0131$         & \citenum{Gaia21aa}                   \\          
TESS\ Input Catalog designation           & TIC 237277760               & \citenum{Stassun+19aj}               \\
$B$ magnitude                             & $9.638 \pm 0.023$           & \citenum{Hog+00aa}                   \\          
$V$ magnitude                             & $9.183 \pm 0.022$           & \citenum{Hog+00aa}                   \\          
$J$ magnitude                             & $8.268 \pm 0.023$           & \citenum{Cutri+03book}               \\
$H$ magnitude                             & $8.070 \pm 0.017$           & \citenum{Cutri+03book}               \\
$K_s$ magnitude                           & $8.027 \pm 0.022$           & \citenum{Cutri+03book}               \\
Spectral type                             & F3\,V + F3\,V               & \citenum{Popper71apj2}, \citenum{Milone+05aa} \\[3pt]       
\end{tabular}
\end{table}



\targ\ was found to be eclipsing by Strohmeier in 1959 \cite{Strohmeier59vebam,Strohmeier62ibvs} with a period of 1.682\,d and a name of BV\,241. Fitzgerald \cite{Fitzgerald64pddo} obtained 19 photographic spectra with a reciprocal dispersion of 33\,\AA\,mm$^{-1}$, measured radial velocities (RVs) and determined minimum masses for the stars. He found the spectral lines of the two components to be of approximately equal strength, and the orbital period to be double that given by Strohmeier because the primary and secondary eclipses were practically indistinguishable.


Popper \cite{Popper71apj2} obtained 17 photographic spectra of \targ\ at 16.1\,\AA\,mm$^{-1}$, using the Lick 120-inch (3.1\,m) telescope and coud\'e spectrograph. Popper also gave minimum masses, but not the physical properties of the system because no light curve was available. 


A good $V$-band light curve of \targ\ was obtained by Popper \& Dumont \cite{PopperDumont77} and analysed by Popper \& Etzel \cite{PopperEtzel81aj}. They found the stars to have almost identical radii and surface brightnesses, and that ratios of the radii between 0.95 and 1.05 gave nearly indistinguishable fits to the data.


Light curves containing 476 points in each of the $B$ and $V$ bands, and covering most of the eclipse phases, were published by Ibano{\v{g}}lu \etal\ \cite{Ibanoglu+76ibvs} and analysed by G\"ud\"ur \etal\ \cite{Gudur+79aas}. They combined their analysis of these data with the spectroscopic results of Popper \cite{Popper71apj2} to determine the properties of \targ\ for the first time. Russo \etal\ \cite{Russo+81apss} reanalysed the same data with a different approach, and obtained similar results. A comparable dataset was obtained by Chis \etal\ \cite{Chis+80ibvs}, and normal points and an analysis were published by Christesen \etal\ \cite{Cristeson+81aca}.


Milone \etal\ \cite{Milone+05aa} (hereafter M05) presented a measurement of the properties of the system based on light curves in the $H_{\rm P}$, $B_{\rm T}$ and $V_{\rm T}$ bands from the \hip\ satellite and medium-resolution (resolving power $R = 20\,000$) spectroscopy from the Asiago 1.82\,m telescope in the region of the near-infrared calcium triplet. They estimated a spectral type of F3\,V and a sub-solar metallicity of $\FeH \approx -0.4$ from comparisons with spectral atlases. The $B_{\rm T}$ and $V_{\rm T}$ light curves were highly scattered, and the $H_{\rm P}$ photometry included only 12 datapoints during eclipse, so their measurements of the radii of the stars were imprecise. To the author's knowledge, no further analysis of this system has been published.


\section*{Photometric observations}

\begin{figure}[t] \centering \includegraphics[width=\textwidth]{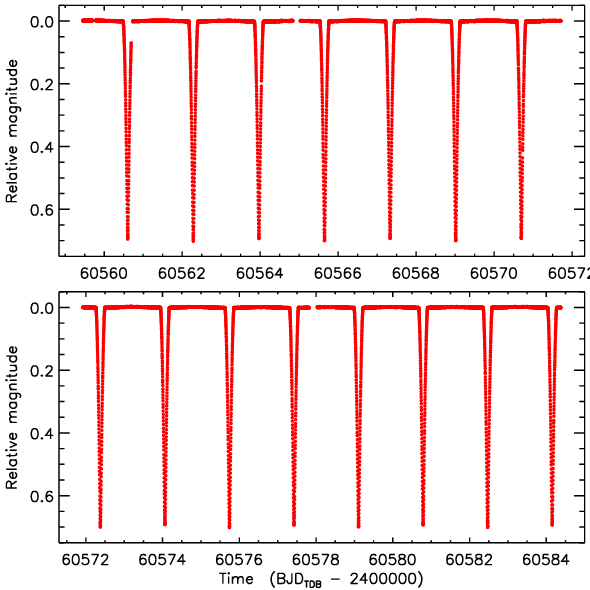} \\
\caption{\label{fig:time} Light curve of \targ\ from TESS sector 83, chosen as it 
has very few data gaps. The flux measurements have been converted to magnitude 
units and the median subtracted.} \end{figure}

\targ\ is the subject of an abundance of data from the TESS mission, which has been used to observe it in a total of 40 sectors. Thirty-eight of these were observed at the best (120-s) cadence, sector 41 was observed at 600-s cadence, and sector 59 at 200-s cadence. We used the {\sc lightkurve} package \cite{Lightkurve18} to download the SPOC (Science Processing Center \cite{Jenkins+16spie}) light curves for all 40 sectors from the NASA Mikulski Archive for Space Telescopes (MAST\footnote{\texttt{https://mast.stsci.edu/portal/Mashup/Clients/Mast/Portal.html}}), specifying the ``hard'' option to reject data flagged as lower quality. 

The data were converted from flux into differential magnitude and the median magnitude was subtracted from each sector. We then visually inspected all sectors to reject stretches of data away from fully-observed transits, ending with a total of 603,334 datapoints from the original 667,710 points (!). A representative plot from sector 83 is shown in Fig.~\ref{fig:time}.

We queried the \gaia\ DR3 database\footnote{\texttt{https://vizier.cds.unistra.fr/viz-bin/VizieR-3?-source=I/355/gaiadr3}} for a list of all sources within 2~arcmin of \targ. This search returned 44 sources in addition to the target itself, the brightest two being 3.4~mag and 3.9~mag fainter in the \Grp\ band than \targ\ itself. We thus expect the amount of third light in the TESS data of our target to be small.

\section*{Light curve analysis}

The profusion of TESS data available for \targ\ demands the use of a fast modelling code for its interpretation. As the two stars are well-separated, we used version 44 of the {\sc jktebop}\footnote{\texttt{http://www.astro.keele.ac.uk/jkt/codes/jktebop.html}} code \cite{Me++04mn2,Me13aa}. Our analysis followed that for UZ\,Dra in Paper~27 \cite{Me25obs6}. We fitted each TESS sector separately, to allow for changes in parameters such as third light. For each sector we fitted for the orbital period ($P$), a reference time of primary minimum ($T_0$) close to the midpoint of that sector, the sum ($r_{\rm A}+r_{\rm B}$) and ratio ($k = r_{\rm B}/r_{\rm A}$) of the fractional radii, the central surface brightness ratio ($J$), third light ($L_3$), and orbital inclination ($i$). 

\begin{figure}[t] \centering \includegraphics[width=\textwidth]{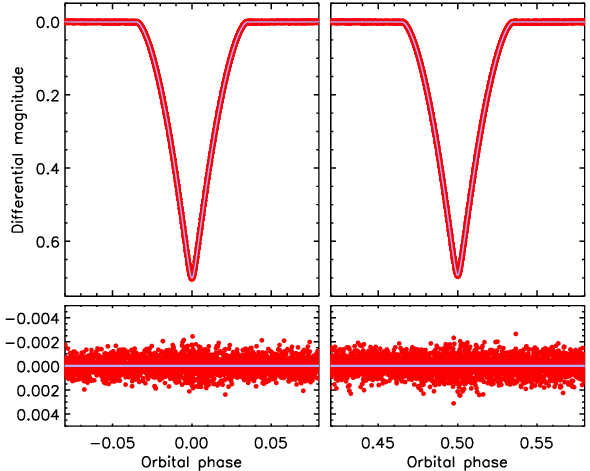} \\
\caption{\label{fig:phase} {\sc jktebop} best fit to the light curves of \targ\ from 
TESS sector 83 for the primary eclipse (left panels) and secondary eclipse (right panels). 
The data are shown as filled red circles and the best fit as a light blue solid line. 
The residuals are shown on an enlarged scale in the lower panels.} \end{figure}

Limb darkening (LD) was allowed for using the power-2 law \cite{Hestroffer97aa,Maxted18aa,Me23obs2}. As the two stars are almost identical, we adopted the same LD coefficients for both, fitted for the linear coefficient ($c$) and fixed the non-linear coefficient ($\alpha$) at a theoretical value \cite{ClaretSouthworth22aa,ClaretSouthworth23aa}. We also fitted for several nuisance parameters: the coefficients of a straight line representing the out-of-eclipse brightness of the system for each chunk of data, and the coefficients of the reflection effect for both stars. We assumed the orbit to be circular as there is no indication of eccentricity in the available data. An example fit can be found in Fig.~\ref{fig:phase}.

As an experiment, we initially treated sector 41, which has a cadence of 600~s, identically to the other sectors. The results showed good agreement with the other sectors with the exception of the orbital inclination, which was lower by 0.05$^\circ$ ($5\sigma$). We then accounted for the lower cadence by numerically integrating the model in the fitting process \cite{Me11mn} and found that the results shifted to complete agreement with other sectors. Sector 59 was observed at 200-s cadence: the fitted parameters are in full agreement with other sectors even without accounting for the slightly lower sampling rate.

The final photometric parameters were calculated by taking the unweighted mean and standard deviation of the values from the individual sectors, and are given in Table~\ref{tab:jktebop}. We refrained from adopting the standard errors of the parameters because they are too small -- the agreement between sectors is so good that the standard errors are significantly smaller than the limit to which we trust our modelling code (see ref.~\citenum{Maxted+20mn} for further information). We also calculated uncertainties for each sector using Monte Carlo (MC) simulations \cite{Me08mn} (after scaling the data errors to give a reduced \chisq\ of $\chir = 1$), finding that the MC errorbars slightly underpredict the scatter between the parameter values for different sectors. 

\begin{table} \centering
\caption{\em \label{tab:jktebop} Photometric parameters measured using 
{\sc jktebop} from the TESS light curves of \targ. The errorbars are 
the standard deviation of the results for individual sectors.}
\begin{tabular}{lcc}
{\em Parameter}                           &              {\em Value}            \\[3pt]
{\it Fitted parameters:} \\
Orbital inclination (\degr)               & $      89.487      \pm  0.010     $ \\
Sum of the fractional radii               & $       0.21652    \pm  0.00008   $ \\
Ratio of the radii                        & $       0.9934     \pm  0.0010    $ \\
Central surface brightness ratio          & $       0.9935     \pm  0.0008    $ \\
Third light                               & $       0.0099     \pm  0.0036    $ \\
LD coefficient $c$                        & $       0.602      \pm  0.010     $ \\
LD coefficient $\alpha$                   &              0.4984 (fixed)         \\
{\it Derived parameters:} \\
Fractional radius of star~A               & $       0.10861    \pm  0.00007   $ \\       
Fractional radius of star~B               & $       0.10791    \pm  0.00007   $ \\       
Light ratio $\ell_{\rm B}/\ell_{\rm A}$   & $       0.9808     \pm  0.0017    $ \\[3pt]
\end{tabular}
\end{table}

\begin{figure}[t] \centering \includegraphics[width=\textwidth]{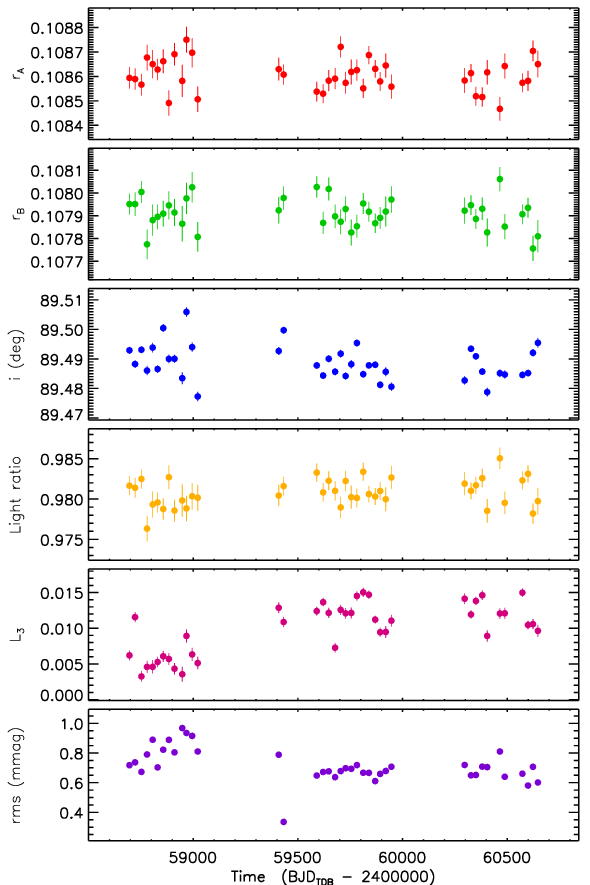} \\
\caption{\label{fig:parget} The best fit to selected photometric parameters
of \targ\ from all TESS sectors. The times used on the abscissae are given in 
Table~\ref{tab:tmin}. The errorbars are from the MC simulations.} \end{figure}

Fig.~\ref{fig:parget} shows the values of some of photometric parameters for each TESS sector. We see the same story as UZ\,Dra: that the astrophysical parameters are very stable with time but third light varies slightly. This is expected because TESS has a coarse pixel scale and the satellite's orientation changes each sector, so background stars may drift in and out of the pixel mask used to generate the photometry of our target star. We also see that the light ratio is definitively less than unity which, combined with the small but clear difference in depth of the two eclipses in Fig.~\ref{fig:phase}, means our identification of the hotter star (star~A) is certain.


\newpage

\section*{Orbital ephemeris}

\begin{table} \centering
\caption{\em Times of mid-eclipse for \targ\ and their residuals versus the fitted ephemeris. \label{tab:tmin}}
\setlength{\tabcolsep}{10pt}
\begin{tabular}{rrrrr}
{\em Orbital} & {\em Eclipse time}  & {\em Uncertainty} & {\em Residual} & {\em TESS}   \\
{\em cycle}   & {\em (BJD$_{TDB}$)} & {\em (d)}         & {\em (d)}      & {\em sector} \\[3pt]
$-307.0$ & 2458695.259746 & 0.000003 & $ 0.000005$ & 14 \\   
$-299.0$ & 2458722.171840 & 0.000003 & $ 0.000000$ & 15 \\   
$-290.0$ & 2458752.447953 & 0.000003 & $ 0.000003$ & 16 \\   
$-282.0$ & 2458779.360048 & 0.000003 & $-0.000000$ & 17 \\   
$-274.0$ & 2458806.272136 & 0.000004 & $-0.000010$ & 18 \\   
$-267.0$ & 2458829.820231 & 0.000003 & $-0.000001$ & 19 \\   
$-259.0$ & 2458856.732334 & 0.000003 & $ 0.000004$ & 20 \\   
$-251.0$ & 2458883.644433 & 0.000003 & $ 0.000004$ & 21 \\   
$-243.0$ & 2458910.556525 & 0.000003 & $-0.000002$ & 22 \\   
$-232.0$ & 2458947.560666 & 0.000005 & $ 0.000004$ & 23 \\   
$-226.0$ & 2458967.744729 & 0.000004 & $-0.000006$ & 24 \\   
$-218.0$ & 2458994.656834 & 0.000004 & $ 0.000000$ & 25 \\   
$-210.0$ & 2459021.568933 & 0.000003 & $ 0.000001$ & 26 \\   
$ -95.0$ & 2459408.430344 & 0.000003 & $ 0.000001$ & 40 \\   
$ -88.0$ & 2459431.978433 & 0.000003 & $ 0.000004$ & 41 \\   
$ -41.0$ & 2459590.087008 & 0.000003 & $ 0.000002$ & 47 \\   
$ -32.0$ & 2459620.363120 & 0.000003 & $ 0.000004$ & 48 \\   
$ -24.0$ & 2459647.275209 & 0.000003 & $-0.000006$ & 49 \\   
$ -15.0$ & 2459677.551331 & 0.000003 & $ 0.000006$ & 50 \\   
$  -7.0$ & 2459704.463425 & 0.000003 & $ 0.000002$ & 51 \\   
$   0.0$ & 2459728.011504 & 0.000003 & $-0.000005$ & 52 \\   
$   8.0$ & 2459754.923605 & 0.000003 & $-0.000002$ & 53 \\   
$  16.0$ & 2459781.835703 & 0.000003 & $-0.000003$ & 54 \\   
$  25.0$ & 2459812.111814 & 0.000002 & $-0.000002$ & 55 \\   
$  33.0$ & 2459839.023909 & 0.000002 & $-0.000005$ & 56 \\   
$  42.0$ & 2459869.300024 & 0.000002 & $-0.000001$ & 57 \\   
$  49.0$ & 2459892.848112 & 0.000003 & $ 0.000001$ & 58 \\   
$  57.0$ & 2459919.760204 & 0.000004 & $-0.000005$ & 59 \\   
$  65.0$ & 2459946.672303 & 0.000003 & $-0.000004$ & 60 \\   
$ 169.0$ & 2460296.529590 & 0.000003 & $ 0.000007$ & 73 \\   
$ 178.0$ & 2460326.805698 & 0.000002 & $ 0.000004$ & 74 \\   
$ 185.0$ & 2460350.353776 & 0.000002 & $-0.000004$ & 75 \\   
$ 194.0$ & 2460380.629888 & 0.000003 & $-0.000002$ & 76 \\   
$ 201.0$ & 2460404.177971 & 0.000003 & $-0.000005$ & 77 \\   
$ 219.0$ & 2460464.730197 & 0.000003 & $-0.000000$ & 79 \\   
$ 226.0$ & 2460488.278293 & 0.000003 & $ 0.000010$ & 80 \\   
$ 251.0$ & 2460572.378596 & 0.000002 & $ 0.000006$ & 83 \\   
$ 259.0$ & 2460599.290683 & 0.000002 & $-0.000005$ & 84 \\   
$ 266.0$ & 2460622.838778 & 0.000003 & $ 0.000004$ & 85 \\   
$ 273.0$ & 2460646.386858 & 0.000005 & $-0.000002$ & 86 \\   
\end{tabular}
\end{table}

\begin{figure}[t] \centering \includegraphics[width=\textwidth]{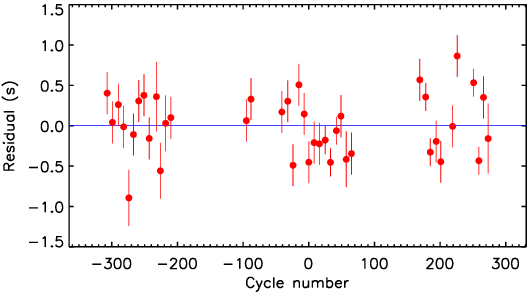} \\
\caption{\label{fig:tmin} Residuals of the times of minimum light from 
Table~\ref{tab:tmin} (red circles) versus the best-fitting ephemerides. 
The blue solid line indicates a residual of zero. The ordinate axis does 
indeed show only $\pm$1.5 seconds.} \end{figure}

The analysis above returned a measurement of the mean time of primary eclipse for each TESS sector. We fitted a linear ephemeris to these times, obtaining
\begin{equation}
\mbox{Min~I} = {\rm BJD}_{\rm TDB}~ 2459728.011509 (2) + 3.364012273 (4) E
\end{equation}
where $E$ is the number of cycles since the reference time of minimum and the bracketed quantities indicate the uncertainty in the final digit of the previous number. The scatter around the best fit is larger than the errorbars suggest, with $\chir = 2.2$, so the uncertainties in the ephemeris have been multiplied by $\sqrt{\chir}$ to account for this. The rms scatter is a remarkable 0.37~s: \targ\ has deep V-shaped eclipses which are optimal for the precise determination of the time of midpoint. The individual timings are given in Table~\ref{tab:tmin} and the residuals are plotted in Fig.~\ref{fig:tmin}. 

We also fitted quadratic and cubic functions of time to compare to the linear ephemeris, finding that they do not give an improved fit. We set a $3\sigma$ upper limit of $9.6 \times 10^{-11}$~s~s$^{-1}$ (2.0~ms~yr$^{-1}$) on the rate of change of orbital period. This limit could be improved if published times of minimum light were added to the analysis, but this is outside the scope of the current work.



\section*{Radial velocity analysis}

Three previous studies of \targ\ have presented RVs: Fitzgerald \cite{Fitzgerald64pddo} obtained 19 coud\'e photographic spectra; Popper \cite{Popper71apj2} observed 17 coud\'e photographic spectra with twice the dispersion; and M05 presented RVs from 27 \'echelle spectra obtained with a CCD. We have copied all the RVs from these works and refitted them using {\sc jktebop} to check the results (Table~\ref{tab:sb}), adopting a circular orbit. \reff{The $P$ was fixed to 3.364012273~d, and $T_0$ and the velocity amplitudes $K_{\rm A}$ and $K_{\rm B}$ were the fitted parameters.} We used 1000 Monte Carlo simulations each to 

\clearpage

\noindent obtain uncertainties \cite{Me21obs5}. Transformation to a standard scale is not accounted for in the values or uncertainties of the systemic velocities.

The Fitzgerald RVs are provided with weights. We converted these into uncertainties and scaled them to obtain $\chir=1.0$ for the RVs for each star individually. We then obtained two fits to the RVs for the two stars: with one systemic velocity for the system ($V_{\rm \gamma}$) and with separate systemic velocities for the two stars ($V_{\rm \gamma,A}$ and $V_{\rm \gamma,B}$). Our results agree with those of Fitzgerald's to within the errorbars. We find a larger rms scatter, partly because we quote the value for all RVs whereas Fitzgerald's value was for RVs with unit weight.

\begin{sidewaystable} \centering
\caption{\em Spectroscopic orbits for \targ\ from the literature and from the current work. 
In each case two sets of orbits are given: where the systemic velocity for the two stars are 
forced to be the same or allowed to differ. The adopted result is based on all RVs and 
different systemic velocities. \reff{The $K_{\rm A}$ and $K_{\rm B}$ values for M05 were calculated
from other parameters given in that work.} All quantities are given in km~s$^{-1}$. \label{tab:sb}}
\setlength{\tabcolsep}{10pt}
\begin{tabular}{lccccccc}
{\em Source} & {\em $K_{\rm A}$} & {\em $K_{\rm B}$} & {\em $V_{\rm \gamma}$} & {\em $V_{\rm \gamma,A}$} & {\em $V_{\rm \gamma,B}$} & {\em  $\sigma_{\rm A}$} & {\em $\sigma_{\rm B}$} \\[10pt]

Fitzgerald \cite{Fitzgerald64pddo}  & $ 99.4 \pm 1.7$ & $100.4 \pm 1.7$ & $+1.3 \pm 1.0$ &                &                & 2.0 & 2.0 \\
This work (Fitzgerald RVs)          & $100.1 \pm 1.1$ & $101.1 \pm 1.8$ &                & $+0.4 \pm 1.0$ & $+2.3 \pm 1.4$ & 5.3 & 6.7 \\
This work (Fitzgerald RVs)          & $100.0 \pm 1.1$ & $100.9 \pm 1.8$ & $+1.0 \pm 0.8$ &                &                & 5.2 & 6.7 \\[5pt]

Popper \cite{Popper71apj2}          & $ 99.1 \pm 0.9$ & $ 99.2 \pm 1.2$ &                & $+2.2 \pm 0.8$ & $+0.4 \pm 1.0$ & 3.1 & 4.1 \\
This work (Popper RVs)              & $ 99.3 \pm 1.1$ & $ 99.2 \pm 0.7$ &                & $+0.0 \pm 1.0$ & $+2.2 \pm 0.7$ & 3.8 & 2.8 \\
This work (Popper RVs)              & $ 99.3 \pm 1.2$ & $ 99.2 \pm 0.8$ & $+1.6 \pm 0.6$ &                &                & 4.1 & 2.9 \\[5pt]

M05                                 & \reff{(96.6)}   & \reff{(98.0)}   & $-0.5 \pm 1.5$ &                &                & 0.19 & 0.15 \\
This work (M05 RVs, adopted)        & $ 96.8 \pm 0.6$ & $ 98.4 \pm 0.5$ &                & $-0.6 \pm 0.4$ & $-0.5 \pm 0.4$ & 2.34 & 2.05 \\
This work (M05 RVs)                 & $ 96.8 \pm 0.6$ & $ 98.4 \pm 0.5$ & $-0.6 \pm 0.3$ &                &                & 2.34 & 2.05 \\
\end{tabular}
\end{sidewaystable} 



The Popper RVs are not provided with weights so we assumed an equal uncertainty for each star: that which gave $\chir=1.0$ (see Table~\ref{tab:sb}). Our identification of which is the primary star differs from Popper's so we swapped the RV datafiles for the two stars. Our results are in good agreement with those of Popper, including the finding of a slightly different systemic velocity for the two stars (with a significance of $1.8\sigma$). 

The RVs from M05 were not provided with individual uncertainties so we again assigned the same uncertainty to all RVs per star in order to obtain $\chir=1.0$. Our finding of 2.3 and 2.1\kms, for star~A and star~B respectively, is much greater than the 0.19 and 0.15\kms\ quoted by M05 (their table~4). M05 also did not fit spectroscopic orbits but instead included their RVs with the \hip\ light curve in a global fit which directly yielded mass and radius measurements. \reff{We have calculated the equivalent $K_{\rm A}$ and $K_{\rm B}$ values and inserted them in Table~\ref{tab:sb}.}

\begin{figure}[t] \centering \includegraphics[width=\textwidth]{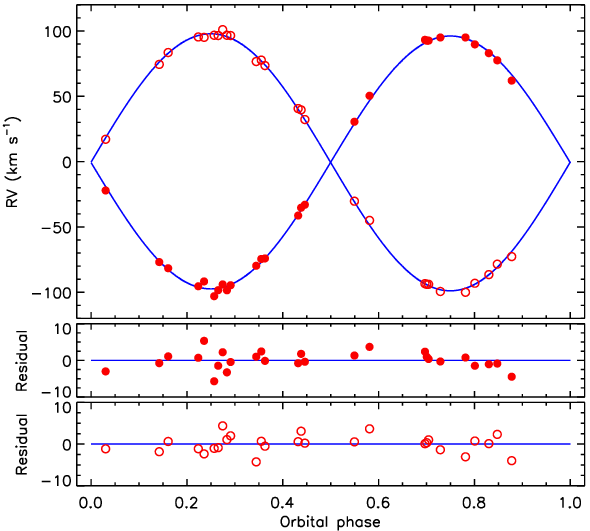} \\
\caption{\label{fig:rv} RVs of \targ\ from M05 compared to the best fit from 
{\sc jktebop} (solid blue lines). The RVs for star~A are shown with filled 
circles, and for star~B with open circles. The residuals are given in the 
lower panels separately for the two components.} \end{figure}

The two photographic RV datasets both give values of $K_{\rm A}$ and $K_{\rm B}$ which are larger than the modern equivalents (M05 RVs) and have slight discrepancies between the systemic velocities for the two stars. Because of this, and that the M05 RVs are both more numerous and more precise than either photographic study, we adopt our fit to the M05 RVs (with separate systemic velocities) as the spectroscopic orbit of \targ\ in the following analysis. A plot of the RVs against our fits is in Fig.~\ref{fig:rv}. \reff{If we had instead taken the weighted mean of all three $K_{\rm A}$ and $K_{\rm B}$ values we would have obtained $97.9 \pm 0.7$ and $98.8 \pm 0.6$\kms. In the near future new RVs from \gaia\ DR4 will be available for checking the $K_{\rm A}$ and $K_{\rm B}$ values for this system.}

\section*{Physical properties and distance to \targ}

We used the $r_{\rm A}$, $r_{\rm B}$ and $i$ values from Table~\ref{tab:jktebop}, the orbital period from the ephemeris given above, and the $K_{\rm A}$ and $K_{\rm B}$ values from Table~\ref{tab:sb} to determine the physical properties of the \targ\ system. The calculations were performed using standard formulae \cite{Hilditch01book} implemented in the {\sc jktabsdim} code \cite{Me++05aa}, which propagates uncertainties from the input to the output parameters by perturbation. The results are given in Table~\ref{tab:absdim}.

M05 gave the effective temperature (\Teff) of our star~A (which is their secondary star) as $6626 \pm 153$~K from spectral fitting. Our $J$ from Table~\ref{tab:jktebop} corresponds to a \Teff\ difference of 11~K. Putting these measurements together and rounding to 10~K gives the values we adopted (Table~\ref{tab:absdim}).

\begin{table} \centering
\caption{\em Physical properties of \targ\ defined using the nominal solar units 
given by IAU 2015 Resolution B3 (ref.~\citenum{Prsa+16aj}). \label{tab:absdim}}
\begin{tabular}{lr@{~$\pm$~}lr@{~$\pm$~}l}
{\em Parameter}        & \multicolumn{2}{c}{\em Star A} & \multicolumn{2}{c}{\em Star B}    \\[3pt]
Mass ratio   $M_{\rm B}/M_{\rm A}$          & \multicolumn{4}{c}{$0.9837 \pm 0.0077$}       \\
Semimajor axis of relative orbit (\Rsunnom) & \multicolumn{4}{c}{$12.974 \pm 0.051$}        \\
Mass (\Msunnom)                             &  1.305  & 0.015       &  1.284  & 0.017       \\
Radius (\Rsunnom)                           &  1.4092 & 0.0056      &  1.4001 & 0.0055      \\
Surface gravity ($\log$[cgs])               &  4.2559 & 0.0022      &  4.2544 & 0.0027      \\
Density ($\!\!$\rhosun)                     &  0.4665 & 0.0020      &  0.4679 & 0.0021      \\
Synchronous rotational velocity ($\!\!$\kms)& 21.19   & 0.08        & 21.06   & 0.08        \\
Effective temperature (K)                   &  6630   & 150         &  6620   & 150         \\
Luminosity $\log(L/\Lsunnom)$               &  0.539  & 0.039       &  0.530  & 0.040       \\
$M_{\rm bol}$ (mag)                         &  3.39   & 0.10        &  3.41   & 0.10        \\
Interstellar reddening \EBV\ (mag)          & \multicolumn{4}{c}{$0.05 \pm 0.02$}	        \\
Distance (pc)                               & \multicolumn{4}{c}{$191.3 \pm 2.9$}           \\[3pt]
\end{tabular}
\end{table}



Our mass and radius measurements have precisions of 1.3\% and 0.4\%, respectively, which is significantly better than previous analyses have given \cite{Gudur+79aas,Russo+81apss,Milone+05aa} and the first time all values have been obtained to 2\% precision. \reff{We note, however, that adopting the weighted-mean velocity amplitudes from the end of the previous section would increase the values of the masses by 2.1\% for star~A and 2.8\% for star~B. The forthcoming Data Release 4 from the \gaia\ satellite will provide new RVs with which to check these values.}


To estimate the distance to the system we used the $BV$ magnitudes from Tycho \cite{Hog+00aa}, the $JHK_s$ magnitudes from 2MASS \cite{Cutri+03book} (corrected to the Johnson system using transformations from Carpenter \cite{Carpenter01aj}), the stars' physical properties from Table~\ref{tab:absdim}, and calibrations of surface brightness versus \Teff\ from Kervella et al.\ \cite{Kervella+04aa}. The optical and infrared distance estimates agree with the inclusion of an interstellar reddening of $\EBV = 0.05 \pm 0.02$~mag. The best distance estimate is $191.3 \pm 2.9$~pc based on the $K_{\rm s}$ band, which is in acceptable agreement with the parallax distance of $195.34 \pm 0.50$~pc from \gaia\ DR3 \cite{Gaia23aa}. 


We estimated the age of the system by comparing the measured masses, radii and \Teff s to theoretical predictions from the {\sc parsec} 1.2 evolutionary models \cite{Bressan+12mn}. The best fit occurs for an age of $1600 \pm 150$~Myr and a fractional metal abundance by mass of $Z=0.014$. An age of $1700 \pm 150$~Myr matches the masses and radii for $Z=0.017$, but the predicted \Teff\ values are slightly low. No solution can be found that matches the radii and \Teff\ values of the components, for their measured masses, for $Z=0.020$ or $Z=0.010$. These results agree with the suggestion by M05 that \targ\ is mildly metal-poor.



\section*{Stellar activity}


\begin{figure}[t] \centering \includegraphics[width=\textwidth]{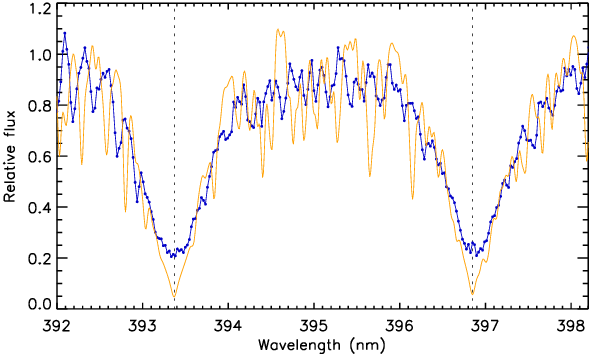} \\
\caption{\label{fig:cahk} Observed spectrum of \targ\ around the Ca~{\sc ii} H 
and K lines (blue line with points) compared to a synthetic spectrum for a star 
with $\Teff = 6630$~K, $\logg = 4.0$ and solar metallicity from the BT-Settl model 
atmospheres \cite{Allard+01apj,Allard++12rspta} (orange line). The H and K line 
central wavelengths are shown with dotted lines. The spectrum of \targ\ has been 
shifted to zero velocity and normalised to approximately unit flux.} \end{figure}

We obtained one spectrum of the Ca~{\sc ii} H and K lines of \targ\ on the night of 2022/06/07 to search for emission lines due to chromospheric activity. We used the Isaac Newton Telescope (INT) and Intermediate Dispersion Spectrograph (IDS), with the 235~mm camera, H2400B grating, EEV10 CCD, a 1~arcsec slit, and an exposure time of 300~s. The data were reduced using a pipeline currently being written by the author \cite{Me+20mn2}, which performs bias subtraction, division by a flat-field from a tungsten lamp, aperture extraction, and wavelength calibration using copper-argon and copper-neon arc lamp spectra. The spectrum has a resolution of approximately 0.05~nm, a reciprocal dispersion of 0.023~nm~px$^{-1}$, a coverage of 373--438~nm, and a signal-to-noise ratio of 45. 

The observation was obtained at an orbital phase of 0.905, when the RV separation of the stars was 109\kms\ (0.144~nm). Fig.~\ref{fig:cahk} shows the spectrum compared to a synthetic spectrum of the same atmospheric parameters from the BT-Settl model atmospheres \cite{Allard+01apj,Allard++12rspta}. The Ca~{\sc ii} H and K lines show some infilling as expected from chromospheric emission, but no variations due to starspots are seen in the TESS light curves. This suggests that \targ\ has chromospheric activity but negligible starspot activity.

\section*{Summary and conclusions}

We have presented an analysis of the dEB \targ, which contains two almost-identical F3 stars in a circular orbit of period 3.36~d. We have determined the masses of the stars to 1.3\% and their radii to 0.4\%, using new light curves from 40 sectors of the TESS mission and published spectroscopic results. For the first time we clearly detect a difference in depth between the primary and secondary eclipses, allowing a definitive assignation of which is the primary star. We showed that it is slightly but significantly hotter, larger and more massive than its companion. The distance we found to the system agrees with the \gaia\ DR3 parallax.

The eclipses are deep and triangular in shape, a morphology that is optimal for measuring precise eclipse times. We determined one overall time of primary eclipse per TESS sector, from all primary and secondary eclipses in that sector. The measurements have a remarkably low scatter of 0.37~s around the best-fitting linear ephemeris, indicating that \targ\ may be useful in checking the timings of future datasets. This has already been done for TESS by Von Essen \etal\ \cite{Vonessen+20aj}, who found the satellite's timings to be $5.8 \pm 2.5$~s \reff{earlier} than ground-based observations of 26 binaries showing deep eclipses. Similar dEBs may be useful in cross-checking timings of the TESS and forthcoming PLATO \cite{Rauer+25exa} missions.

By comparing the physical properties of \targ\ to theoretical models we deduced an age of $1600 \pm 300$~Myr and a slightly sub-solar metallicity. More precise \Teff\ measurements would be helping in refining the age; a spectroscopic metallicity measurement would also permit the reliability of the models to be assessed. 

The stars are hot enough to show $\delta$~Scuti or $\gamma$~Doradus pulsations, and such a discovery \reff{would be} scientifically valuable \cite{MeVanreeth22mn,MeBowman26araa}. We thus checked for any pulsations in the system by calculating the frequency spectrum of the residuals of the {\sc jktebop} fit to sectors 47--60, using version 1.2.0 of the {\sc period04} code \cite{LenzBreger05coast}. None were found in the frequency interval 0--100~d$^{-1}$ to a limit of $10^{-5}$~mag.


\section*{Acknowledgements}

We thank the anonymous referee for a prompt, positive and helpful report.
This paper includes data collected by the TESS\ mission and obtained from the MAST data archive at the Space Telescope Science Institute (STScI). Funding for the TESS\ mission is provided by the NASA's Science Mission Directorate. STScI is operated by the Association of Universities for Research in Astronomy, Inc., under NASA contract NAS 5–26555.
This paper includes observations made with the Isaac Newton Telescope operated on the island of La Palma by the Isaac Newton Group of Telescopes in the Spanish Observatorio del Roque de los Muchachos of the Instituto de Astrof\'{\i}sica de Canarias.
This work has made use of data from the European Space Agency (ESA) mission {\it Gaia}\footnote{\texttt{https://www.cosmos.esa.int/gaia}}, processed by the {\it Gaia} Data Processing and Analysis Consortium (DPAC\footnote{\texttt{https://www.cosmos.esa.int/web/gaia/dpac/consortium}}). Funding for the DPAC has been provided by national institutions, in particular the institutions participating in the {\it Gaia} Multilateral Agreement.
The following resources were used in the course of this work: the NASA Astrophysics Data System; the SIMBAD database operated at CDS, Strasbourg, France; and the ar$\chi$iv scientific paper preprint service operated by Cornell University.



\begin{thebibliography}{10}
\newcommand{\enquote}[1]{`#1'}

\bibitem{Andersen91aarv}
J.~{Andersen}, \textit{A\&ARv}, \textbf{3}, 91, 1991.

\bibitem{Torres++10aarv}
G.~{Torres}, J.~{Andersen} \& A.~{Gim{\'e}nez}, \textit{A\&ARv}, \textbf{18},
  67, 2010.

\bibitem{Maxted+20mn}
P.~F.~L. {Maxted} \textit{et~al.}, \textit{MNRAS}, \textbf{498}, 332, 2020.

\bibitem{Me15aspc}
J.~{Southworth}, in \textit{Living Together: Planets, Host Stars and Binaries}
  (S.~M. {Rucinski}, G.~{Torres} \& M.~{Zejda}, eds.), 2015,
  \textit{Astronomical Society of the Pacific Conference Series}, vol. 496, p.
  321.

\bibitem{Me20obs}
J.~{Southworth}, \textit{The Observatory}, \textbf{140}, 247, 2020.

\bibitem{Me21univ}
J.~{Southworth}, \textit{Universe}, \textbf{7}, 369, 2021.

\bibitem{Ricker+15jatis}
G.~R. {Ricker} \textit{et~al.}, \textit{Journal of Astronomical Telescopes,
  Instruments, and Systems}, \textbf{1}, 014003, 2015.

\bibitem{Borucki16rpph}
W.~J. {Borucki}, \textit{Reports on Progress in Physics}, \textbf{79}, 036901,
  2016.

\bibitem{Hog+00aa}
E.~{H{\o}g} \textit{et~al.}, \textit{A\&A}, \textbf{355}, L27, 2000.

\bibitem{Cutri+03book}
R.~M. {Cutri} \textit{et~al.}, \textit{{2MASS All Sky Catalogue of Point
  Sources}} (The IRSA 2MASS All-Sky Point Source Catalogue, NASA/IPAC Infrared
  Science Archive, Caltech, US), 2003.

\bibitem{Gaia23aa}
{Gaia Collaboration}, \textit{A\&A}, \textbf{674}, A1, 2023.

\bibitem{CannonPickering20anhar}
A.~J. {Cannon} \& E.~C. {Pickering}, \textit{Annals of Harvard College
  Observatory}, \textbf{95}, 1, 1920.

\bibitem{ESA97}
ESA, \textit{ESA Special Publication}, \textbf{1200}, 1997.

\bibitem{Gaia21aa}
{Gaia Collaboration}, \textit{A\&A}, \textbf{649}, A1, 2021.

\bibitem{Stassun+19aj}
K.~G. {Stassun} \textit{et~al.}, \textit{AJ}, \textbf{158}, 138, 2019.

\bibitem{Popper71apj2}
D.~M. {Popper}, \textit{ApJ}, \textbf{166}, 361, 1971.

\bibitem{Milone+05aa}
E.~F. {Milone} \textit{et~al.}, \textit{A\&A}, \textbf{441}, 605, 2005.

\bibitem{Strohmeier59vebam}
W.~{Strohmeier}, \textit{Veroeffentlichungen der Remeis-Sternwarte zu Bamberg},
  \textbf{5}, 3, 1959.

\bibitem{Strohmeier62ibvs}
W.~{Strohmeier}, \textit{IBVS}, \textbf{9}, 1, 1962.

\bibitem{Fitzgerald64pddo}
M.~P. {Fitzgerald}, \textit{Publications of the David Dunlap Observatory},
  \textbf{2}, 417, 1964.

\bibitem{PopperDumont77}
D.~M. {Popper} \& P.~J. {Dumont}, \textit{AJ}, \textbf{82}, 216, 1977.

\bibitem{PopperEtzel81aj}
D.~M. {Popper} \& P.~B. {Etzel}, \textit{AJ}, \textbf{86}, 102, 1981.

\bibitem{Ibanoglu+76ibvs}
C.~{Ibano{\v{g}}lu} \textit{et~al.}, \textit{IBVS}, \textbf{1100}, 1, 1976.

\bibitem{Gudur+79aas}
N.~{G\"ud\"ur} \textit{et~al.}, \textit{A\&AS}, \textbf{36}, 65, 1979.

\bibitem{Russo+81apss}
G.~{Russo} \textit{et~al.}, \textit{Ap\&SS}, \textbf{79}, 359, 1981.

\bibitem{Chis+80ibvs}
D.~{Chis} \textit{et~al.}, \textit{IBVS}, \textbf{1794}, 1, 1980.

\bibitem{Cristeson+81aca}
C.~{Cristeson} \textit{et~al.}, \textit{AcA}, \textbf{31}, 505, 1981.

\bibitem{Lightkurve18}
{Lightkurve Collaboration}, \enquote{{Lightkurve: Kepler and TESS time series
  analysis in Python}}, Astrophysics Source Code Library, 2018.

\bibitem{Jenkins+16spie}
J.~M. {Jenkins} \textit{et~al.}, in \textit{Proc.\ SPIE}, 2016, \textit{Society
  of Photo-Optical Instrumentation Engineers (SPIE) Conference Series}, vol.
  9913, p. 99133E.

\bibitem{Me++04mn2}
J.~{Southworth}, P.~F.~L. {Maxted} \& B.~{Smalley}, \textit{MNRAS},
  \textbf{351}, 1277, 2004.

\bibitem{Me13aa}
J.~{Southworth}, \textit{A\&A}, \textbf{557}, A119, 2013.

\bibitem{Me25obs6}
J.~{Southworth}, \textit{The Observatory}, \textbf{145}, 201, 2025.

\bibitem{Hestroffer97aa}
D.~{Hestroffer}, \textit{A\&A}, \textbf{327}, 199, 1997.

\bibitem{Maxted18aa}
P.~F.~L. {Maxted}, \textit{A\&A}, \textbf{616}, A39, 2018.

\bibitem{Me23obs2}
J.~{Southworth}, \textit{The Observatory}, \textbf{143}, 71, 2023.

\bibitem{ClaretSouthworth22aa}
A.~{Claret} \& J.~{Southworth}, \textit{A\&A}, \textbf{664}, A128, 2022.

\bibitem{ClaretSouthworth23aa}
A.~{Claret} \& J.~{Southworth}, \textit{A\&A}, \textbf{674}, A63, 2023.

\bibitem{Me11mn}
J.~{Southworth}, \textit{MNRAS}, \textbf{417}, 2166, 2011.

\bibitem{Me08mn}
J.~{Southworth}, \textit{MNRAS}, \textbf{386}, 1644, 2008.

\bibitem{Me21obs5}
J.~{Southworth}, \textit{The Observatory}, \textbf{141}, 234, 2021.

\bibitem{Hilditch01book}
R.~W. {Hilditch}, \textit{{An Introduction to Close Binary Stars}} (Cambridge
  University Press, Cambridge, UK), 2001.

\bibitem{Me++05aa}
J.~{Southworth}, P.~F.~L. {Maxted} \& B.~{Smalley}, \textit{A\&A},
  \textbf{429}, 645, 2005.

\bibitem{Prsa+16aj}
A.~{Pr{\v s}a} \textit{et~al.}, \textit{AJ}, \textbf{152}, 41, 2016.

\bibitem{Carpenter01aj}
J.~M. {Carpenter}, \textit{AJ}, \textbf{121}, 2851, 2001.

\bibitem{Kervella+04aa}
P.~{Kervella} \textit{et~al.}, \textit{A\&A}, \textbf{426}, 297, 2004.

\bibitem{Bressan+12mn}
A.~{Bressan} \textit{et~al.}, \textit{MNRAS}, \textbf{427}, 127, 2012.

\bibitem{Allard+01apj}
F.~{Allard} \textit{et~al.}, \textit{ApJ}, \textbf{556}, 357, 2001.

\bibitem{Allard++12rspta}
F.~{Allard}, D.~{Homeier} \& B.~{Freytag}, \textit{Philosophical Transactions
  of the Royal Society of London Series A}, \textbf{370}, 2765, 2012.

\bibitem{Me+20mn2}
J.~{Southworth} \textit{et~al.}, \textit{MNRAS}, \textbf{497}, 4416, 2020.

\bibitem{Vonessen+20aj}
C.~{von Essen} \textit{et~al.}, \textit{AJ}, \textbf{160}, 34, 2020.

\bibitem{Rauer+25exa}
H.~{Rauer} \textit{et~al.}, \textit{Experimental Astronomy}, \textbf{59}, 26,
  2025.

\bibitem{MeVanreeth22mn}
J.~{Southworth} \& T.~{Van Reeth}, \textit{MNRAS}, \textbf{515}, 2755, 2022.

\bibitem{MeBowman26araa}
J.~{Southworth} \& D.~{Bowman}, \textit{ARA\&A, in press,
  \texttt{arXiv:2509.08426}}, 2026.

\bibitem{LenzBreger05coast}
P.~{Lenz} \& M.~{Breger}, \textit{Communications in Asteroseismology},
  \textbf{146}, 53, 2005.

\end{thebibliography}

\end{document}